# Active role of the nonmagnetic cations in magnetic interactions for the double-perovskite $Sr_2BOsO_6$ (B=Y, In, Sc).

Sudipta Kanungo,[1,*] Binghai Yan,[1,2,3] Claudia Felser,[1] and Martin Jansen [1,4]

[1] Max-Planck-Institut für Chemische Physik fester Stoffe, 01187, Dresden, Germany

[2] School of Physical Science and Technology, Shanghai Tech University, Shanghai 200031, China

[3] Max-Planck-Institut für Physik komplexer Systeme, 01187, Dresden, Germany

[4] Max-Planck-Institut für Festkörperforschung, 70569, Stuttgart, Germany

Using first-principles density functional theory, we have investigated the electronic and magnetic properties of recently synthesized and characterized $4d$-$5d$ double-perovskites $Sr_2BOsO_6$ (B=Y, In, Sc). The electronic structure calculations show that in all compounds, the $Os^{5+}$ ($5d^3$) site is the only magnetically active one, while $Y^{3+}$, $In^{3+}$ and $Sc^{3+}$ remain in nonmagnetic states, with Sc/Y and In featuring $d^0$ and $d^{10}$ electronic configurations, respectively. Our studies reveal the important role of closed-shell ($d^{10}$) versus open-shell ($d^0$) electronic configurations of the nonmagnetic sites in determining the overall magnetic exchange interactions. Although the magnetic $Os^{5+}$ ($5d^3$) site is the same in all compounds, the magnetic super-exchange interactions mediated by non-magnetic Y/In/Sc species are strongest for $Sr_2ScOsO_6$, weakest for $Sr_2InOsO_6$, and intermediate in case of the Y ($d^0$), due to different energy overlaps between Os-$5d$ and Y/In/Sc-$d$ states. This explains the experimentally observed substantial differences in the magnetic transition temperatures of these materials, despite of an identical magnetic site and underlying magnetic ground state. Furthermore, short range Os-Os exchange-interactions are more prominent than long range Os-Os interactions in these compounds, which contrasts with the behavior of other $3d$-$5d$ double-perovskites.

**Introduction**

Double-perovskite oxides with the general formula $A_2BB'O_6$ are extensively studied and have attracted enormous attention from material science communities over decades owing to their diverse physical properties such as colossal magneto-resistance (e.g., $Sr_2FeMoO_6$[1]), multiferroicity (e.g., $Ba_2NiMnO_6$[2]), room temperature magneto-dielectric properties (e.g., $La_2NiMnO_6$[3]), optical properties (e.g., $Sr_2CrReO_6$[4]), high $T_N$ ferrimagnetism (e.g., $Sr_2CrOsO_6$[5]), ferromagnetic-insulator (e.g., $Ca_2CrSbO_6$[6]), half metallicity (e.g., $A_2CrWO_6$[7]) and metallicity (e.g., $Sr_2CrReO_6$[8]). The key elements dictating these properties are transition metals with different $d$ orbitals in the B and B' sub-lattices. Individually the B and B' sub-lattices form two fcc lattices. The entire structure is a combination of two interpenetrating fcc lattices, each of which exhibits intrinsic geometric frustration. The wide range of choices for the B and B' ions, provide a great tunability of intra- and inter-sublattice interactions. The subtle competition between different exchange interactions can lead to exotic magnetic phases such as, antiferromagnetic (AFM) transitions in the 3d-5d double-perovskites $Sr_2CoOsO_6$[9,10] and $Sr_2FeOsO_6$[11,12]. Here, the long-range Os-Co/Fe-Os coupling is found to be surprisingly large mediated by the magnetic Co ($3d^7$)/Fe ($3d^5$).

To investigate the general trend of long-range $5d$-$5d$ exchange coupling, it was considered that $d^0$ and $d^{10}$ configurations at the B site represent two extreme cases for non-magnetic cations. Recently compounds $Sr_2ScOsO_6$ (SSOO), $Sr_2YOsO_6$ (SYOO), and $Sr_2InOsO_6$ (SIOO) with the AFM transition temperatures $T_N$=92, 53, and 27 K respectively have been synthesized [13,14]. They display a surprisingly large range of magnetic ordering temperatures, which obviously depend on the type of the non-magnetic B cation incorporated [13,14]. To rationalize the causes of such unexpected behavior, we address these observations in the present theoretical study. The effect of electron filling on magnetic properties has been discussed for other materials in the previous literature[15,16].

In this work, we focus on SYOO and SIOO, both of which have a nonmagnetic B-4$d$ site and a magnetic B'-5$d$ site [$Os^{5+}(d^3)$], to investigate the effect of the nonmagnetic cations on the Os-Os exchange couplings. For SYOO, the nonmagnetic site $Y^{3+}$ has a $4d^0$ open shell configuration. In SIOO, $In^{3+}$ has a closed shell $4d^{10}$ configuration. These ions provide an ideal platform to compare the $d^0$ and $d^{10}$ cases. In addition, SSOO is similar to SYOO in that, the only difference is the presence of the 3$d$ $Sc^{3+}$ ion at the nonmagnetic site, which offers an opportunity to explore 4$d$ versus 3$d$ cases. We have performed electronic structure calculations based on density-functional theory (DFT) and have analyzed the short-range and long-range magnetic exchange interactions of the Os sub-lattice. Our calculations reveal that the short-range Os-Os interactions are much stronger than the long-range ones, unlike other double-perovskites such as $Sr_2CoOsO_6$[9,10] and $Sr_2FeOsO_6$[11,12]. We find that the exchange coupling depends strongly on the overlap between the Os-5$d$ and Y/In/Sc-$d$ states in same energy window. The hybridization between Os-5$d$ and In-4$d$ is much smaller in the $d^{10}$ close-shell case than that in the $d^0$ open-shell case, which results in the amplitudes of Os-Os coupling being smaller in the In compound than in the Y and Sc compounds. The smaller exchange coupling in the Y compound versus Sc is attributed to a similar origin, wherein the hybridization of Sc-3$d$ states with the Os-5$d$ states in the density of states (DOS) is greater than that for Y-4$d$. Thus, we can successfully explain the varied trend in the Neel temperatures observed experimentally.

**Crystal Structure and Computational Details:**

SYOO, SIOO and SSOO crystallize in a monoclinic structure with space group P2$_1$/n. The theoretically optimized structures are obtained by relaxing the atomic positions of all atoms, while keeping the lattice parameters fixed at their experimentally determined low-temperature (2.9 K) values[13]. The structures consist of alternating corners sharing distorted $BO_6$ (B=Y, In, Sc) and $OsO_6$ octahedron with Sr atoms situated at the void positions between the two types of octahedron. The six metal-oxygen bond lengths of the distorted octahedra are grouped into three different values. Because of the monoclinic distortion, the in-plane and out-of-plane B-O-Os (B=Y, In, Sc) chains deviate substantially (<B-O-Os ~155º-160º) from an ideal 180º values as shown in Table I. The DFT calculations are performed with the plane-wave basis set based on a pseudo-potential framework, as implemented in the Vienna *ab-initio* simulation package (VASP)[17]. The generalized gradient approximation (GGA) exchange-correlation functional is employed following the Perdew-Burke-Ernzerhof prescription[18] for calculating the electronic structures. Further the effect of correlation on the electronic structure was investigated by performing additional GGA+U[19,20] calculations. For the plane-wave basis, a 600 eV plane-wave cutoff is applied and a k-point mesh of 8 × 8 × 6 in the Brillouin zone is used for the self-consistent calculations.

Table I. Experimental crystal structure data for SYOO, SIOO and SSOO.

|  | SYOO | SIOO | SSOO |
|---|---|---|---|
| Volume (Å$^3$) | 274.8 | 260.8 | 253.9 |
| Bond lengths (Os-O) (Å) | 1.94, 1.96, 1.99 | 1.94, 1.97, 1.95 | 1.95, 1.96, 1.96 |
| Bond Angles (<O-Os-O) | 90.2, 90.9, 90.5 | 92.5, 92.7, 90.9 | 92.6, 90.7, 90.8 |
| Bond Angles (<B-O-Os) | 157.7, 156.6, 155.2 (c) | 160.6, 153.9, 160.5 (c) | 165.7, 166.2, 166.3 (c) |

**Results and Discussions**

First we investigate the electronic structure within GGA prescription and the calculated ferromagnetic (FM) DOS is shown in Fig.1. The top, middle and bottom panels represent the SYOO, SIOO and SSOO DOS, respectively. The Sr states lie far above the Fermi level ($E_f$) and are not shown in the figure, which is consistent with the nominal $Sr^{2+}$ valence state. Note that the spin-polarized calculations within the GGA, without an artificial Coulomb *U*, drive the insulating solution with very small gap at $E_f$. For each compound, only the Os states contribute to the $E_f$ along with a substantial portion from the O-2$p$ states, while all Y/In/Sc-$d$ states are either completely empty or filled. For SYOO, the Y-4$d$ states are com-pletely empty and lie almost 5 eV above the $E_f$, confirming the presence of the $Y^{3+}$ state with a $4d^0$ configuration. On the other hand, in SIOO, the In-4$d$ states are almost 10 eV below the $E_f$, (not shown in the figure), consistent with a completely filled $In^{3+}$ $4d^{10}$ shell. The Os-5$d$ states split according to the octahedral environment of the surrounding oxygen atoms, with the $t_{2g}$ states being completely filled in the majority spin channel and completely empty in the minority spin channel. The Os-$e_g$ states are completely empty in both channels. The calculated GGA magnetic moments at the Os site are 2.03, 1.99, and 1.97 $\mu_B$ for SYOO, SIOO and SSOO respectively, and are consistent with the experimentally measured effective magnetic moments [13]. The Y/In/Sc sites remain non-magnetic with zero magnetic moment. The magnetic moment at the O sites are also non-negligible [0.115 $\mu_B$ (SYOO), 0.123 $\mu_B$ (SIOO), 0.123 $\mu_B$ (SSOO)] reflecting the strong hybridization between O-2$p$ and Os-5$d$ states. Together with the calculated magnetic moments, these findings suggest that Os is in the 5+ ($5d^3$) valence electronic state with a high spin (S=3/2) configuration. This result implies that for these compounds, Os is the only magnetically active site in these compounds with combinations of $d^0$-$d^3$ (SYOO, SSOO) and $d^{10}$-$d^3$ (SIOO) configurations. Thus, an effective spin model can be constructed in terms of only Os-$t_{2g}$ degrees of freedom. To reveal the effect of the Coulomb $U$, on the electronic structure, we did a systematic analysis of electronic structure with $U_{eff}^{Os}$= 1 to 3 eV[14,27], (shown in supplementary[35] Fig.1 and Table. I) and found that, a systematic increment in the band gap and magnetic moments at the Os site take place, however no significant changes occurred in the electronic structure, spin states of the compounds in the entire range of $U$ compared to GGA calculations.

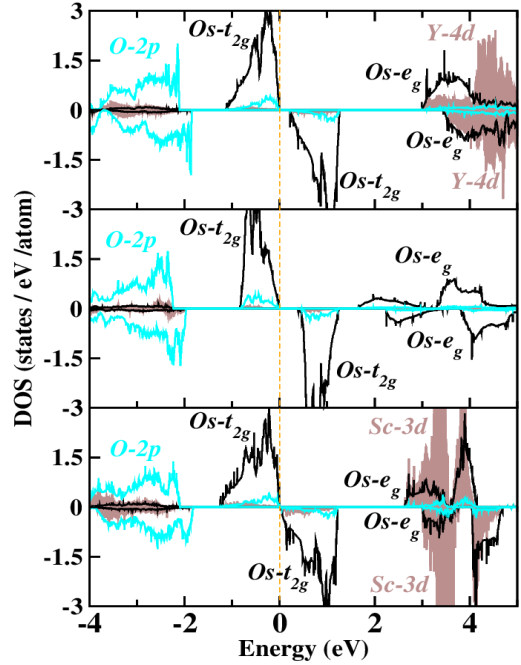

Fig.1 (Color online) GGA FM density of states. The top to bottom panels show the DOS for SYOO, SIOO and SSOO, projected onto the Y-4$d$--Os-5$d$, the In-4$d$--Os-5$d$ and Sc-3$d$--Os-5$d$ states respectively, along with the O-2$p$ states. The two channels for each panel represent majority and minority spin channels. The $E_f$ is marked at zero on the energy scale.

Experimental measurements [13] show that the AFM transition temperatures ($T_N$) of these compounds are very different. To understand the trend in $T_N$, we decided to calculate the magnetic exchange interactions with DFT based first-principles calculations. Exchange interactions can be calculated using the Kugel-Khomskii model [22,23], which requires the correct choice of $U$, the Hunds exchange, and a proper estimate of charge transfer energies between different orbitals. However, because of the complex exchange paths involving different types of atoms and orbitals, such energies are difficult to estimate. Therefore, we chose a different route using the total energy calculations of various spin configurations, and then mapped the DFT total energies onto the corresponding Ising models[24] with the equation $E^{total} = \Sigma_{ij} J_{ij} \sigma_i \sigma_j$, where $J_{ij}$ is the magnetic exchange interaction between the $i^{th}$ and $j^{th}$ sites and the $\sigma's$ are the effective spin values corresponding to the respective sites. Although this method suffers from several drawbacks, such as the choice of spin configurations, and exchange-correlation functional, the method has helped in successfully estimating qualitative trends in exchange interactions for various types of materials[9-11,25,26]. We used both the GGA and GGA+$U$ ($U_{eff}^{Os}$=1 and 3 eV[14,27]) to estimate the exchange couplings. To probe the long-range exchange interaction, we created a 2 x 2 x 2 super-cell and considered five independent exchange pathways connecting various Os sites, as illustrated in Fig.2 (a). The point to be noted that GGA and GGA+$U$ calculation gave qualitatively same trend of exchange interactions for the three compounds with a slight decrease in absolute values due to inclusion of the $U$. The GGA results are listed in Table II, while GGA+$U$ results are tabulated in supplementary material[35] Table III. We also cross-checked the convergence of J's values upon varying calculation parameters and found that it depends less than 1% on the calculation parameters such as plane wave cutoff, k-points, energy convergence and no of bands.

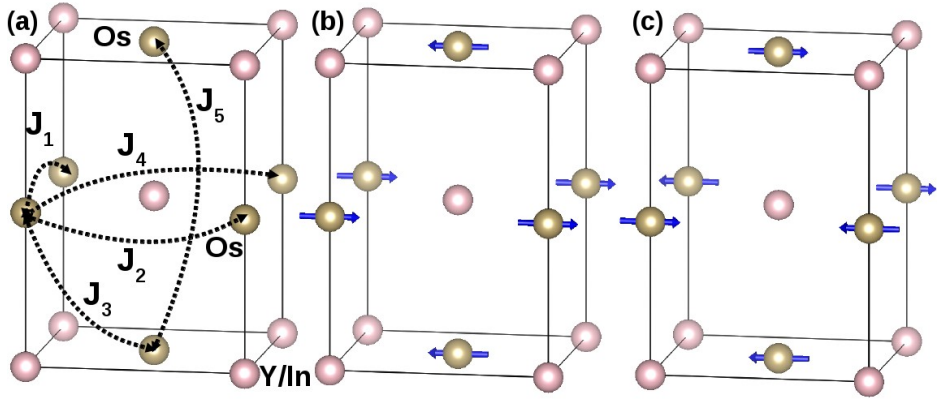

Fig. 2. (Color online) (a) Magnetic exchange paths connecting different Os sites (brown spheres) in the monoclinic unit cell of $Sr_2BOsO_6$ (B=Y, In). Sr and O atoms are omitted from the structure for clarity. The five interaction paths between different Os sites are denoted by $J_1$ to $J_5$. The two lowest magnetic spin structures; (b) experimental spin structure AFM-I (c) 2$^{nd}$ lowest spin structure AFM-II.

Two interesting trends can be identified from the results. First, the short range Os-Os interactions are much larger than the long range Os-Os interactions. For example, the Os-O-B-O-Os nearest-neighbor interactions ($J_1$, $J_2$, and $J_3$) are much stronger than the next-nearest neighbor interactions ($J_4$, and $J_5$). This trend is opposite to what we found in previous studies of 3$d$-5$d$ double-perovskite $Sr_2CoOsO_6$[9,10], $Sr_2FeOsO_6$[11], and $Sr_2NiIrO_6$[28], where the nearest-neighbor interactions are much weaker than the next-nearest one. The major electronic difference is that, in the present case Os is the only magnetically active site, whereas for the compounds in the previous studies[9,10,11,28] both B (Fe, Co, Ni) and B' (Os, Ir) were magnetically active.

Table II: Magnetic exchange interactions calculated with GGA for the paths, shown in Fig. 2(a). The values of J (meV) for SYOO, SIOO and SSOO are listed in the table below with (+) and (-) signs indicating AFM and FM interactions respectively. Exchange interaction values for $Sr_2FeOsO_6$ are shown for comparison from Ref. 11.

| Interaction Paths [Os-Os] | $Sr_2YOsO_6$ | $Sr_2InOsO_6$ | $Sr_2ScOsO_6$ | $Sr_2FeOsO_6$ [11] |
|---|---|---|---|---|
| $J_1$ (in plane short range) | 3.02 | 1.16 | 5.24 | - 0.2 |
| $J_2$ (in plane short range) | 4.63 | 2.52 | 7.42 | |
| $J_3$ (out of plane short range) | 3.90 | 1.47 | 6.43 | 3.3 |
| $J_4$ (in plane long range) | 0.91 | 0.04 | 1.66 | - 6.8 |
| $J_5$ (out of plane long range) | 0.76 | 0.08 | 0.42 | 12.8 |

Because the transition metal at site B is magnetically inactive in the present compounds, the long-range super-super-exchange interaction between two Os sites (Os-O-B-O-Os), connected by a 180° B site mediated ligand network, becomes weaker than the direct nearest neighbor direct interactions between two Os sites connected by a 90° ligand network. This is clearly visible in the Fig.3.

Another interesting observation is that the strength of exchange-interactions is stronger for SYOO than for SIOO. For example, the strongest nearest-neighbor interactions ($J_1$, $J_2$, and $J_3$) for SYOO are almost double those of SIOO, while the next-nearest-neighbor interactions are an order of magnitude stronger for SYOO than for SIOO. These interesting differences can be visualized using a localized Wannier function representation. Fig.3 shows the plots of effective Wannier-like orbitals located at Os sites corresponding to the Os-Os interactions for both SYOO (top panel) and SIOO (bottom panel). The central parts comprised Os-$d$ characters, while the tails situated at the different sites are shaped according to the integrated out orbitals. The weights at the neighboring tails dictate the strength of interactions between different sites. From Fig.3a, we can see that the $J_5$ (Os-O-Y-O-Os) interaction is very small compared to the $J_2$ and $J_3$ interactions, because there are large $d$ tails at the connecting Os sites for $J_2$ and $J_3$, relative to $J_5$. Fig.3b shows that, the in-plane nearest-neighbor $J_1$ and $J_2$ (Os-O-Y-O-Os) 90° interactions are similar in strength, as indicated by the similar weight of the connecting Os tails, which are much greater than for $J_4$. It has been suggested in previous studies[9-12, 28] that long-range 5$d$-5$d$ interactions are much stronger than the short range one due to the extended nature of 5$d$ wave functions. However, in the present case even though we have a magnetically active 5$d$ site, the short-range Os-Os interactions ($J_1$, $J_2$, and $J_3$) are much stronger than the long-range one ($J_4$, $J_5$), because of the presence of an interpenetrating nonmagnetic rather than magnetic B sublattice.

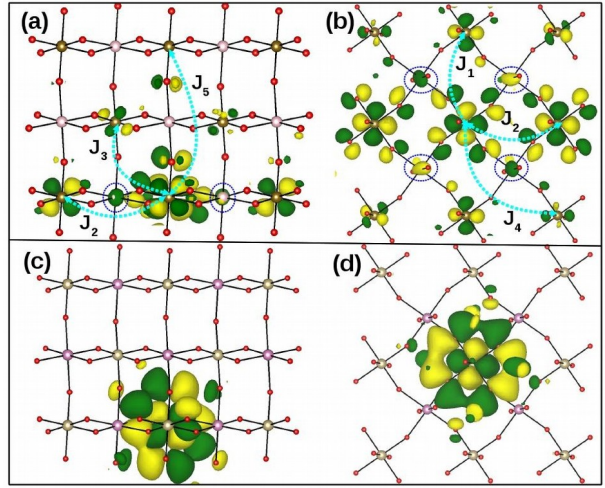

Fig.3. (Color online) The Wannier functions of Os-$t_{2g}$'s for SYOO and SIOO calculated with GGA prescription are shown in the top ((a)-(b)) and bottom ((c)-(d)) panels, respectively. The central part of the Os Wannier functions comprises Os-5$d_{yz}$ and Os-5$d_{xy}$ characters for (a,c) and (b,d), respectively. The exchange interactions $J_2$-$J_3$-$J_5$ and $J_1$-$J_2$-$J_4$ are shown in (a) and (b), respectively. Green and yellow colors represent surfaces with isovalues of 0.18 and −0.18, respectively. Red spheres represent O atoms, while the other color symbols are the same as those in Fig.2.

Interestingly, both top panels show significant tails corresponding to the nearest neighbor interaction at the connecting Os sites, while the bottom panel shows almost no tails at the connecting sites. We also observed strong Os-$t_{2g}$ tails at the neighboring Y sites, which are marked by dotted circles in Fig.3(a,b); these are completely absent in the case of In, as shown in Fig.3(c,d). The pictorial representations of the Wannier orbitals clearly indicate the possible role of interpenetrating nonmagnetic Y or In sub-lattices in determining the overall magnetic interactions.

A point to be noted within the GGA and GGA+$U$ description, experimentally observed AFM-I state is energetically higher than AFM-II, as shown in Figs. 2(b) and (c) respectively, however the inclusion of spin-orbit coupling (SOC) with the GGA drove the experimentally observed AFM-I to a value ~1.2 meV/f.u lower than the AFM-II. These results indicate a strong influence of SOC on the electronic structure, as has been suggested in recent studies[29-31] of $4d^3$ and $5d^3$ $t_{2g}$ systems. We calculated the exchange interactions including SOC for the three compounds and found that the relative strengths of the AFM interactions ($J_3$) for SYOO (3.67 meV), SIOO (1.46 meV) and SSOO (5.55 meV) remained unchanged upon addition of SOC. Here we want to mention that, we restrict our self in evaluating only symmetric part of the exchange interactions, ignoring the anti-symmetric ($S_i$ x $S_j$) type of exchanges even in the presence of SOC. The anti-symmetric exchange will only arise due to canting of spins ($S_i$ x $S_j$) and in the collinear spin configuration, this part vanishes. We neglected this anti-symmetric type of exchanges because, the experimentally observed antiferromagnetic configuration is a collinear AFM, where spins are aligned in parallel to the *ab* plane. Therefore, in-principle antisymmetric contribution would be negligible and to make a simplified description to capture the experimental observation, we ignore virtually negligible such complex exchanges in the present case. Moreover the calculated exchange interactions using a symmetric type exchange in the presence of SOC are able to reproduce the experimental trend transition temperatures in this series of compounds. Calculated magneto-crystalline energies (~1-1.5 meV/f.u) favors in-plane spin alignment consistent with the experimental observations. A point to be noted for all three compounds is that Os exhibits rather large orbital moment (-0.11 $\mu_B$), oppositely aligned to that of the spin moments, which is expected for less than half-filled configurations. Although for half-filled $t_{2g}$ case, orbital moments should be quenched, large orbital moments may arise from strong mixing with O-2$p$ states as mentioned in the literature[21].

From our calculations we found that SOC is really important to understand the magnetism in these Os based double-perovskite. We found significant orbital moment at the Os site, which is counter intuitive considering the half-filled $t_{2g}$ orbitals. The effective moment is substantially smaller than what would expect from spin-only value for S=3/2 (3.89 $\mu_B$) state and that only can be explained by taking both positive spin moment and negative orbital moment taken into account. Therefore, our results, go with, results presented in Ref. [31] in context of $Ba_2YOsO_6$, Ref. [32, 33] in context of $Sr_2CrOsO_6$ and contrary to the result presented in Ref. [34] in the context of $Sr_2CrOsO_6$. In the Ref. [34], the authors did not find any evidence of orbital moment at the Os site and based on that they claimed that SOC is not responsible for the finite net moment. Moreover, in the present series, we found that experimentally observed AFM structure can only be stabilized with the inclusion of SOC. Therefore, our results suggest that SOC is important to understand the magnetic behaviors in these materials, however, only symmetric type of exchange is enough to understand the basic experimental observations.

Because $Os^{5+}(5d^3)$, is the only magnetically active site in the two iso-structural, iso-electronic and iso-valent compounds, a common expectation is that they should show similar magnetic interactions. Moreover, it is generally believed that nonmagnetic cations do not play any active role in determining magnetic interactions and transition temperatures. Our studies show opposite scenario. For SYOO and SIOO, $Os^{5+}$ has a $5d^3$ configuration, whereas $Y^{3+}$ and $In^{3+}$ have $4d^0$ and $4d^{10}$ valence configurations, respectively. The super-exchange interactions between the two Os sites are mediated by non-magnetic oxygen and Y or In states. Because $Y^{3+}$ has an open $4d$ shell, it is strongly hybridized with the empty Os-$e_g$ states, as revealed by large Os-$d$ tails at the Y sites in the Wannier plots (marked by circles) and the energy overlap in the DOS. This allow the hopping of Os electrons via empty Y-$4d$ orbitals. In the case of $In^{3+}$, the closed shell $d^{10}$ configuration, which is deep in energy scale, does not allow the Os electrons hopping. To generalize this mechanism of hybridization driven enhancement of magnetic interactions beyond the present two $4d$-$5d$ compounds, we cross-checked our scheme with the $3d$-$5d$ system in SSOO. The calculated magnetic exchange-interactions of SSOO are that of SYOO. The largest exchange-interaction in SSOO is largest among these three compounds and is readily explained by the very strong hybridization between empty Os-$e_g$ states and Sc-$3d$ states, which overlap over almost the entire energy range as shown in the bottom panel of Fig.1. To compare these findings with the experimental results, we calculated the mean field transition temperatures for SSOO, SYOO, and SIOO using the calculated magnetic exchange-interactions. We found that the calculated ratio of mean-field transition temperature $T_N^{YOs}/T_N^{InOs}$ ~ 2.64 (GGA), 2.90 (GGA+U), 2.51 (GGA+SOC) and $T_N^{ScOs}/T_N^{YOs}$ ~1.64 (GGA), 1.66 (GGA+U), 1.51 (GGA+SOC) agreed reasonably well with the experimental $T_N$ ratio of 2.04 and 1.73 respectively. This analysis shows that even though Y, In and Sc are nonmagnetic, their electronic configuration, either open shell or closed shell, will increase or decrease hybridization with the Os states which dictates the strength of the overall magnetic exchange interactions and affects the magnetic transition temperatures of the materials.

Table III. Magnetic exchange interactions ($J_3$) calculated with GGA for SIOO, SYOO ans SSOO in different unit cell volumes. Diagonal bold entries are the values for each compound within its own structure (from Table. I).

| Magnetic Exchange Interaction ($J_3$) meV | | | |
|---|---|---|---|
| | SIOO volume | SYOO volume | SSOO volume |
| SIOO | **1.47** | 1.22 | -- |
| SYOO | 4.38 | **3.90** | 4.66 |
| SSOO | -- | 5.39 | **6.47** |

A relevant question is whether the observed trend in exchange-interactions is due to differences of orbital hybridization induced by the volume contraction and structural distortion of each compounds. It is evident from Table I, although exact values of the bond lengths and angles differ among the three compounds, that there are no drastic structural changes. To clarify this point we calculated the exchange-interactions of the three compounds within the unit cell volumes of the other compounds. The results are summarized in Table III. We found that the trend in exchange-interactions remains unchanged, with maximum difference of absolute values only ± 20%, which confirms that electronic configuration induced hybridization, not volume is the principle determining factor.

**Conclusion**

In summary, we did a comparative analysis of magnetic interactions in the double-perovskites, SYOO, SIOO and SSOO, with a single magnetically active Os site using DFT based first-principles calculations. Although it is known that non-magnetic cations may influence magnetic interactions, our studies provide direct evidence of the role played by the electronic configurations of non-magnetic transition metal cations. Our calculations reveal short-range Os-Os interactions are stronger than long-range Os-Os interactions in the present compounds unlike previous 3*d*-5*d* compounds ($Sr_2FeOsO_6$, $Sr_2CoOsO_6$). Although $Y^{3+}/Sc^{3+}$ and $In^{3+}$ are non-magnetic, their electronic configurations, i.e., open shell ($d^0$) or closed shell ($d^{10}$), strongly influence hybridization with the Os states and the strength of exchange coupling. This leads to the largest $T_N$ for SSOO and the smallest for SIOO, which is consistent with both calculation and experimental observations. The hybridization driven mechanism of enhanced magnetic coupling successfully explains the observed trend in $T_N$ for these systems investigated and can be generalized to the other systems. Our investigation also revealed the importance of SOC in terms of symmetric exchange to understand the magnetism in this series of materials. Our studies highlight the importance of understanding single active site $d^3$ materials and open up a direction for further research in the control of $T_N$ for double-perovskite with single and double magnetic sites.

* Electronic address: Sudipta.Kanungo@cpfs.mpg.de